\documentclass[hyper]{prop2015}
\usepackage[english]{babel}
\usepackage[all]{xy}
\usepackage{bbm}

\category{Proceedings}

\keywords{Green--Schwarz sigma models, M-branes, double dimensional reduction, gauge enhancement, rational homotopy theory, parametrised homotopy theory}
\title{Parametrised Homotopy Theory and Gauge Enhancement}

\subtitle{\href{http://www.maths.dur.ac.uk/lms/109/index.html}{LMS/EPSRC Durham Symposium on Higher Structures in M-Theory}}

\author[V. Braunack-Mayer]{Vincent Braunack-Mayer\inst{a,}\footnote{Corresponding author e-mail:~\href{mailto:vincent.braunack-mayer@uni-hamburg.de}{\textsf{vincent.braunack-mayer@uni-hamburg.de}}}}
\address[1]{Fachbereich Mathematik, Universit\"{a}t Hamburg, Bundesstr. 55, 20146 Hamburg, Germany}

\begin{acknowledgements}
V. B.-M. acknowledges the partial support of the RTG 1670 `Mathematics inspired by String Theory and Quantum Field Theory'.
\end{acknowledgements}

\begin{abstract}
  We review a first-principles derivation of Type IIA D-brane charges from M-theory degrees of freedom in the approximation of super rational homotopy theory.
\end{abstract}
\shortabstract

\begin{document}
\maketitle

\vspace{10pt}
 \noindent
In the article \cite{Braunack-Mayer:2018uyy}, we apply two universal constructions in super rational homotopy theory in order to solve the problem of gauge enhancement of super M-branes (recalled below).
We prove that the topological charge structure of fundamental super M-branes produces twisted K-theory charges of Type IIA D-branes under double dimensional reduction when all torsion effects are ignored\footnote{here \emph{torsion} refers to torsion subgroups of abelian groups, rather than the perhaps more familiar (super-)torsion tensor.}. 
Our result is based upon Sati's conjecture \cite{Sati:2013rxa} that the  topological M-brane charge is classified by degree-$4$ cohomotopy, and our result in turn contributes to a growing body of evidence for this conjecture (see \cite{Fiorenza:2013nha,Fiorenza:2015gla,Fiorenza:2016ypo,Fiorenza:2016oki,Fiorenza:2017jqx,Fiorenza:2018ekd,Huerta:2017utu,Huerta:2018xyh}).

\section{Super $p$-branes and cocycles}
We recall that a \emph{fundamental (or Green--Schwarz) $p$-brane}\footnote{as opposed to a supergravity black brane or a boundary CFT D-brane---fundamental branes are \emph{fundamental} in the sense that they generalise the fundamental Green--Schwarz superstring.} propagating in a spacetime $X$ is described by a $(p+1)$-dimensional sigma model defined on the space of maps
\begin{equation}
\Sigma_{p+1} \longrightarrow X~.
\end{equation}
In this schematic, $\Sigma_{p+1}$ is the abstract worldvolume of the $p$-brane; a $(p+1)$-manifold encoding a $p$-dimensional extended object dynamically evolving in time.
For fundamental \emph{super} $p$-branes propagating on supergravity backgrounds, the worldvolume $\Sigma_{p+1}$ is taken to be a $(p+1)$-dimensional supermanifold.
Compatibility with local supersymmetry on the background $X$ enforces a strict topological constraint: the super volume form on $\Sigma_{p+1}$ must locally trivialise a non-trivial $(p+2)$-cocycle $\mu_{p+2}$ in the supersymmetry super Lie algebra cohomology of super-spacetime \cite{DeAzcarraga:1989vh}.
In this manner, fundamental super $p$-branes are in one-to-one correspondence with such super $(p+2)$-cocycles.
Analysing such \emph{Green--Schwarz sigma models} from the point of view of their controlling super Lie algebra cocycles, hence from the point of view of \emph{super rational homotopy theory}, provides a powerful mathematical toolbox for elucidating many previously elusive aspects of M-theory.

In the case of $d=11$, $\mathcal{N}=1$ supergravity, the local structure of super-spacetime is controlled by the supertranslation super Lie algebra $\mathbbm{R}^{10,1|\mathbf{32}}$  (a classical treatment using different terminology is in \cite{Castellani:1991et}).
The even piece of this super Lie algebra is spanned by the standard basis $\{e_{a}\}_{a=0}^{10}$ of Minkowski space $\mathbbm{R}^{10,1}$, the odd piece is determined by the irreducible real $\mathrm{Spin}(10,1)$-representation $\mathbf{32}$, and the only non-trivial component of the Lie bracket is the odd-odd superbracket
\begin{subequations}
\begin{align}
\mathbf{32}\otimes \mathbf{32} & \longrightarrow \mathbbm{R}^{10,1}~,
\\
(\psi, \varphi) &\longmapsto \left( \overline{\psi}\Gamma^a \varphi\right) e_a~,
\end{align} 
\end{subequations}
determined by the $\mathrm{Spin}(10,1)$-invariant spinor-to-vector pairing (here, as throughout this article, summation over repeated indices is understood).
The corresponding\linebreak Chevalley--Eilenberg algebra $\mathrm{CE}(\mathbbm{R}^{10,1|\mathbf{32}})$  is the super differential graded algebra with underlying algebra structure free on the even degree-$1$ generators $\{e^a\}_{a=0}^{10}$ and odd degree-$1$ generators $\{\psi^\alpha\}_{\alpha=1}^{32}$, corresponding to dual bases of $\mathbbm{R}^{10,1}$ and $\mathbf{32}$ respectively.
The differential ${\rm d}$ on $\mathrm{CE}(\mathbbm{R}^{10,1|\mathbf{32}})$ is obtained by dualising the super Lie bracket; it is defined on generators by
\begin{equation}
{\rm d}\colon
\begin{cases}
e^a\!\!\!\!\!\!\!\!\! &\longmapsto \overline{\psi}\Gamma^a \psi\\
\psi^\alpha\!\!\!\!\!\!\!\!\! &\longmapsto 0
\end{cases}
\end{equation}
and extended to all of $\mathrm{CE}(\mathbbm{R}^{10,1|\mathbf{32}})$ as a graded derivation.

Under the correspondence between cocycles and\linebreak Green--Schwarz sigma models recalled above, the fundamental M2 and M5-branes correspond respectively to the elements
\begin{subequations}
\begin{align}
\mu_{\rm M2} &= \tfrac{i}{2} \overline{\psi}\Gamma_{a_1 a_2} \psi \wedge e^{a_1}\wedge e^{a_2},
\\
\mu_{\rm M5} &=
\tfrac{1}{5!}
\overline{\psi}\Gamma_{a_1\ldots a_5}\psi \wedge e^{a_1}\wedge \ldots \wedge e^{a_5} 
\end{align}
\end{subequations}
in the algebra $\mathrm{CE}(\mathbbm{R}^{10,1|\mathbf{32}})$.
The exceptional Fierz identities for $\mathrm{Spin}(10,1)$ imply the relation
\begin{equation}
\label{eqn:mbrane}
{\rm d} \mu_{\rm M5} = -\tfrac{1}{2} \mu_{M2}\wedge \mu_{M2}~.
\end{equation}
In these algebraic terms, Sati's conjecture on M-brane charge is reflected in the observation that $\mu_{M2}$ and $\mu_{M5}$ combine, via \eqref{eqn:mbrane}, to define a map of (super) commutative differential graded algebras
\begin{equation}
\mathrm{CE}(\mathbbm{R}^{10,1|\mathbf{32}})
\xleftarrow{\;\;\mu_{\rm M2/M5}\;\;}
\mathfrak{l}S^4~,
\end{equation}
where $\mathfrak{l}S^4$ is the minimal model of the rational homotopy type of the sphere $S^4$ (see \cite{hess2007rational} for a survey of rational homotopy theory).
Viewed through the lens of rational homotopy theory, the combined cocycle $\mu_{M2/M5}$ thus encodes the torsion-free part of a map of (super) homotopy types
\begin{equation}
\mathbbm{R}^{10,1|\mathbf{32}}\xrightarrow{\;\;\mu_{\rm M2/M5}\;\;} S^4~,
\end{equation}
controlling the local cohomological charge structure of fundamental M2 and M5-branes propagating in $11$-di\-mensional supergravity.

\section{Gauge enhancement}
\subsection{Double dimensional reduction \dots}
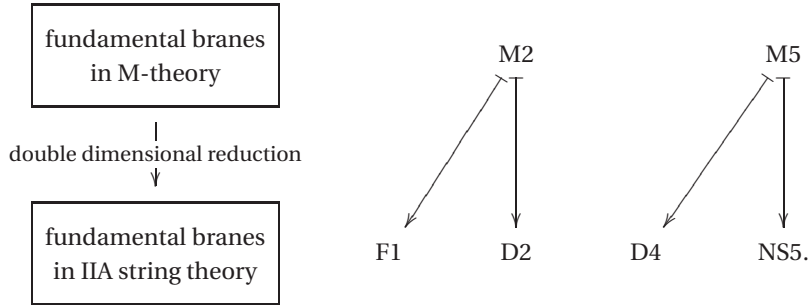
\begin{figure*}
\sidecaption
\hspace{0.16\linewidth}
\begin{equation*}
\raisebox{0pt}{\xymatrix{
\fbox{
\hspace{-4mm}
\begin{tabular}{c}
\small fundamental branes\\
\small in M-theory
\end{tabular}
\hspace{-4mm}
}
\ar[d]|-{\vphantom{\big(}\text{\scriptsize{double dimensional reduction}}}
&
&
\text{M2}
\ar@{|->}[dl]
\ar@{|->}[d]
&
&
\text{M5}\ar@{|->}[dl]
\ar@{|->}[d]
\\
\fbox{
\hspace{-4mm}
\begin{tabular}{c}
\small fundamental branes\\
\small in IIA string theory
\end{tabular}
\hspace{-4mm}
}
&
\text{F1}
&
\text{D2}
&
\text{D4}
&
\text{NS5}.
}}
\end{equation*}
\hspace{0.16\linewidth}
\vspace{-5mm}
\caption{\label{ddrschematic}A schematic of the process of double dimensional reduction, producing Type IIA branes from M-branes.}
\end{figure*}

Insofar as it exists, M-theory is supposed to be the joint non-perturbative completion of perturbative string theory and $11$-dimensional supergravity.
In particular, a complete M-theory must reproduce all of the known structure of perturbative string theory as we pass to various limits.
For instance, dimensional reduction of 11-dimensional supergravity along an $S^1$-fibration yields 10-dimensional Type IIA supergravity equipped with fields corresponding to the Fourier modes in the $S^1$-fibre. In the presence of M2 and M5-branes, the brane worldvolumes either extend along (``wrap'') the $S^1$-fibres or they do not. If, say, the M2-brane worldvolume wraps $S^1$-fibres, then it produces a fundamental string in 10-dimensional Type IIA supergravity; whereas if it does not wrap it produces a D2-brane.
This is the process of \emph{double} dimensional reduction; the dimension of the ambient super-spacetime as well as possibly the brane worldvolume is reduced by one. Figure \ref{ddrschematic} is a schematic of this process.

According to this picture, under double dimensional reduction M-theory readily produces the fundamental string F1, the NS5-brane and the D2 and D4-branes of Type IIA string theory (whereas the D0-brane is also obtained as the Chern class which classifies the 11-dimensional background as an $S^1$-fibration over the 10-dimensional background).
We emphasise that this sche\-matic is a heuristic only; although there are many hints for this double dimensional reduction procedure in the literature \cite{Townsend:1995kk,Townsend:1995af,Figueroa-OFarrill:2002yel,FigueroaO'Farrill:2002tc}, there is no derivation beyond the rational approximation \cite{Fiorenza:2016ypo,Fiorenza:2016oki}.

The problem of gauge enhancement of super M-branes appears within this context of double dimensional reduction.
Perturbative Type IIA string theory admits fundamental $p$-branes in dimensions $p=0,1,2,4,5,6$ and $8$; namely the D0, F1, D2, D4, NS5, D6, and D8-branes respectively.
Moreover, the D$p$-branes are widely expected to carry a combined charge in K-theory twisted by the B-field carried by the fundamental string  \cite{Minasian:1997mm,Witten:9810188,Freed:1999vc,Freed:2000tt,Moore:1999gb}. 
However, the double dimensional reduction schematic (Figure \ref{ddrschematic}) makes no mention of the missing D6 and D8 branes, nor of the combined D-brane charge in twisted K-theory.
The gauge enhancement problem is therefore:
\begin{quote}
How does M-theory give rise to the fundamental D6 and D8 branes, and how does it exhibit the combined D-brane charge in twisted K-theory?
\end{quote}
In \cite{Braunack-Mayer:2018uyy} we provide a full answer to this  question in the torsion-free approximation, using new techniques in super rational homotopy theory from \cite{Braunack-Mayer:2018aa}.
To sketch this solution we first observe that, similarly to our previous discussion, the Green--Schwarz sigma models for fundamental $p$-branes in Type IIA string theory are controlled by elements in the Chevalley--Eilenberg complex $\mathrm{CE}(\mathbbm{R}^{9,1|\mathbf{16}+\overline{\mathbf{16}}})$ of the $d=10$, $\mathcal{N}
=(1,1)$ supertranslation super Lie algebra. Explicitly, these elements are:
\begin{subequations}
\begin{align}
\mu_{\rm F1} & = i(\overline{\psi}\Gamma_a \Gamma_{10}\psi) \wedge e^a~,\\
\mu_{\rm D0} &= \overline{\psi}\Gamma_{10} \psi~,\\
\mu_{\rm D2} &= \tfrac{i}{2} (\overline{\psi} \Gamma_{a_1 a_2} \psi) \wedge e^{a_1}\wedge e^{a_2}~,\\
\mu_{\rm D4}
&= \tfrac{1}{4!} (\overline{\psi} \Gamma_{a_1 \cdots a_4}\Gamma_{10} \psi) \wedge e^{a_1}\wedge\ldots \wedge e^{a_4}~,\\
\mu_{\rm NS5}
&= \tfrac{i}{5!} (\overline{\psi} \Gamma_{a_1 \cdots a_5} \psi) \wedge e^{a_1}\wedge\ldots \wedge e^{a_5}~,\\
\mu_{\rm D6}
&= \tfrac{i}{6!} (\overline{\psi} \Gamma_{a_1 \cdots a_6} \psi) \wedge e^{a_1}\wedge\ldots \wedge e^{a_6}~,\\
\mu_{\rm D8}
&= \tfrac{1}{8!} (\overline{\psi} \Gamma_{a_1 \cdots a_8}\Gamma_{10} \psi) \wedge e^{a_1}\wedge\ldots \wedge e^{a_8}~,
\end{align}
\end{subequations}
controlling the fundamental string, D0, D2, D5, NS5, D6 and D8-branes respectively.
These elements satisfy various compatibility conditions amongst themselves:
\begin{subequations}
\begin{align}
{\rm d}\mu_{\rm F1}&= 0~, \\
{\rm d}\mu_{\rm D0} &= 0~,\\
{\rm d}\mu_{{\rm D}(2p+2)} &= \mu_{\rm F1}\wedge \mu_{{\rm D}(2p)}~, \quad p\in\{0,1,2,3\}~,\\
{\rm d}\mu_{\rm NS5} &= \mu_{\rm D0}\wedge \mu_{\rm D4} - \tfrac{1}{2} \mu_{\rm D2}\wedge \mu_{\rm D2}~.
\end{align}
\end{subequations}
In particular, the F1 and D$p$-brane cocycles together define a cocycle in the rational image of twisted K-theory.

The first universal construction appearing in our solution of the gauge enhancement problem is an adjunction
\begin{equation}
\xymatrix{  
  \mathrm{Spaces}_{/BS^1}
  \ar@{->}@<8pt>[rr]^-{\mathrm{Ext}}
  \ar@{<-}@<-8pt>[rr]^-{\bot}_-{\mathrm{Cyc}}
  &&
  \mathrm{Spaces}
  }
\end{equation}
implementing double dimensional reduction.
Here, the left adjoint $\mathrm{Ext}
$ sends a map of spaces $\tau= (X\to BS^1)$ to the total space of the $S^1$-bundle on $X$ classified by $\tau$ (hence the \emph{extension} of $X$ by the cocycle $\tau$). The  right adjoint $\mathrm{Cyc}$ sends a space $Y$ to the (homotopy) quotient of the free loop space of $Y$ by the action of rigid rotations of loops: $\mathrm{Cyc}(Y) =[S^1, Y]/\!\!/ S^1$.
We implement a version of this adjunction in super rational homotopy theory, where one finds that the $d=11$, $\mathcal{N}=1$ supertranslation super Lie algebra is the central extension of the $d=10$, $\mathcal{N}=(1,1)$ supertranslation super Lie algebra by the D0-brane $2$-cocycle:
\begin{equation}
\mathbbm{R}^{10,1|\mathbf{32}} \cong \mathrm{Ext}\left(\mu_{\rm D0}\colon \mathbbm{R}^{9,1|\mathbf{16}+\overline{\mathbf{16}}}\to BS^1\right)
\end{equation}
(this is an algebraic shadow of \emph{D0-brane condensation}).
Taking the combined M2/M5-brane cocycle, we apply the $\mathrm{Cyc}$ functor and compose with the unit of the $(\mathrm{Ext}\dashv \mathrm{Cyc})$-adjunction to obtain a diagram of super rational homotopy types fibred over $BS^1$:
\begin{equation}
\label{eqn:ddr1stattempt}
\raisebox{24pt}{
\xymatrix{
  \mathbbm{R}^{9,1|\mathbf{16}+\overline{\mathbf{16}}}\ar[d]\ar@/^1pc/[drr]^-{\widetilde{\mu_{\rm M2/M5}}}&&
  \\
  \mathrm{Cyc}\big(\mathbbm{R}^{10,1|\mathbf{32}}\big)\ar[rr]^-{\mathrm{Cyc}(\mu_{\rm M2/M5})}&&
  \mathrm{Cyc}(S^4)~.
  }}
\end{equation}
By direct calculation, we see that the cocycle $\widetilde{\mu_{\rm M2/M5}}$ so obtained reproduces precisely the F1, D2, D4 and NS5-brane cocycles. The D0-brane cocycle is encoded in the map $\mathbbm{R}^{9,1|\mathbf{16}+\overline{\mathbf{16}}}\to BS^1$, hence as the rational image of the Chern character.

\subsection{\dots and gauge enhancement}
In order to complete this picture and produce the D6 and D8-brane cocycles, two additional points are in order.
Firstly, recent results \cite{Huerta:2018xyh} provide an equivariant enhancement of the combined M2/M5-brane cocycle at ADE subgroups of $SU(2)$, establishing a homotopy-theoretic underpinning for the black brane scan. 
The original setting of gauge enhancement of Chan--Paton factors at $\mathbbm{Z}_n$-orbifold points \cite{Witten:1995im} corresponds to  equivariant enhancement of the $\mu_{\rm M2/M5}$ cocycle at A-series subgroups of $SU(2)$.
Considering A-series actions in the limit as $n\to \infty$, the $A_n\sim\mathbbm{Z}_{n+1}$-actions exhaust an $S^1$-action, so that we are naturally led to expect the existence of an $S^1$-equivariant cocycle such as the dotted arrow in the diagram:
\begin{equation}
\xymatrix{
\fbox{
\hspace{-4mm}
\begin{tabular}{c}
\small M-theory\\
\small cocycle
\end{tabular}
\hspace{-4mm}
}
  \ar@{->}[d]|-{\vphantom{\big(}\text{\scriptsize{exhaust $S^1$ fibre}}}
  &
  \mathbbm{R}^{10,1|\mathbf{32}}
  \ar[r]^-{\mu_{\rm M2/M5}}
  \ar[d]
  &
  S^4
  \ar[d]
  \\
\fbox{
\hspace{-4mm}
\begin{tabular}{c}
\small Type IIA\\
\small cocycle
\end{tabular}
\hspace{-4mm}
}
  &\mathbbm{R}^{9,1|\mathbf{16}+\overline{\mathbf{16}}}
  \ar@{-->}[r]^-{\exists?}& S^4/\!\!/S^1.
  }
\end{equation}
However, we find that the existence of such an $S^1$-equi\-variant cocycle is obstructed by nonvanishing of the D4-brane cocycle.

This leads us to our second main point:  homotopical perturbation theory.
Homotopy theory is extremely rich  but very computationally demanding. There is a tower of increasingly accurate approximations to full homotopy theory provided by the \emph{Goodwillie calculus of functors} (see \cite{kuhn2007goodwillie} for a review), which is analogous to analysing smooth functions via their Taylor series expansions.
The first-order approximation of a space $X$ in the Goodwillie calculus is $\Omega^\infty \Sigma^\infty X$, the underlying space of the free infinite loop space on $X$. 
This homotopical linearisation assignment $X\mapsto \Omega^\infty\Sigma^\infty X$ is analogous to ``linearising'' a set $S$ by passing to the free abelian group $\mathbbm{Z}[S]$ and then forgetting the group structure.
There is also such a homotopical linearisation procedure in the relative setting of fibred spaces, where we have an assignment
\begin{equation}
(Y\to X)\longmapsto(\Omega^\infty_X\Sigma^\infty_X Y \to X)
\end{equation}
that applies homotopical linearisation fibrewise to a space $Y$ fibred over a base space $X$.
Working modulo torsion, the main result of the author's thesis \cite{Braunack-Mayer:2018aa} provides algebraic models for parametrised stable rational homotopy types.
In particular, this allows for straightforward algebraic calculations of homotopical linearisations in the torsion-free approximation.

\begin{figure*}
\sidecaption
\hspace{0.1\linewidth}
\begin{equation*}
\xymatrix{
  \mathbbm{R}^{9,1|\mathbf{16}+\overline{\mathbf{16}}}
  \ar[d]
  \ar@{-->}[rrrr]^-{\exists?}
  \ar@/_4pc/[dd]_-{\mu_{\rm F1}}
  \ar@/^1pc/[drr]^>>>>{\widetilde{\mu_{\rm M2/M5}}}
  &&
  &&
  \Omega^\infty_{B^2 U(1)}\Sigma^\infty_{B^2 U(1)} S^4/\!\!/ S^1
  \ar[d]
  \\
  \mathrm{Cyc}(\mathbbm{R}^{10,1|\mathbf{32}})
  \ar[rr]^-{\mathrm{Cyc}(\mu_{\rm M2/M5})}
  \ar[d]
  &&
  \mathrm{Cyc}(S^4)
  \ar@/^1pc/[dll]
  \ar[rr]^-{\text{\scriptsize{fibrewise}}}_-{\text{\scriptsize{linearise}}}
  &&
  \Omega^\infty_{B^2 U(1)}\Sigma^\infty_{B^2 U(1)} \mathrm{Cyc}(S^4)
  \ar@/^1.5pc/[dllll]
  \\
  B^2 U(1)
  &&&&
  }
\end{equation*}
\hspace{0.22\linewidth}
\vspace{-5mm}
\caption{\label{gaugeenhlift}The lifting problem for an $S^1$-equivariant cocycle to first order in homotopical perturbation theory.}
\end{figure*}
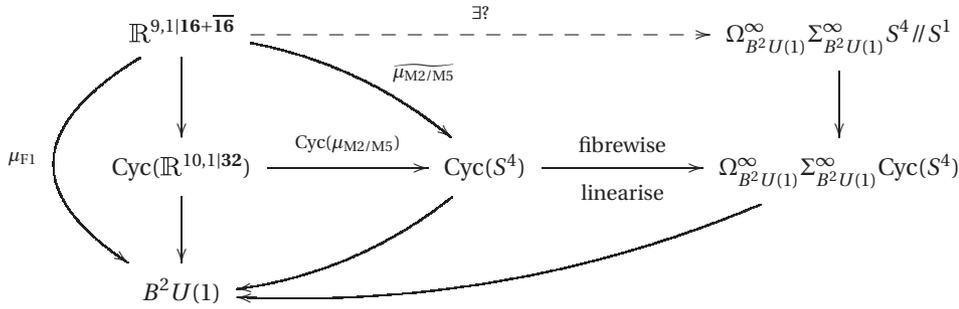

Returning to the gauge enhancement problem, we obtain a solution by passing to homotopical perturbation theory.
We had previously found that passing to the limit over A-series actions does not descend to an $S^1$-equivariant cocycle in $d=10$, $\mathcal{N}=(1,1)$ supergravity.
However, there is a natural comparison map
\begin{equation}
S^4/\!\!/ S^1 \longrightarrow \mathrm{Cyc}(S^4)
\end{equation} 
that fibres rationally over $B^2 U(1)$.
Passing to the fibrewise homotopical linearisations now yields a diagram of super rational homotopy types as in Figure \ref{gaugeenhlift}
and we can ask if the dotted arrow in that diagram exists. This is equivalent to asking whether $\mu_{\rm M2/M5}$ descends to an $S^1$-equivariant cocycle \emph{up to first order} in homotopical perturbation theory.
Using the algebraic models of \cite{Braunack-Mayer:2018aa}, we find that
\begin{enumerate}[i)]
  \item there exists a morphism 
  \begin{equation}
  \widehat{\mu_{\rm M2/M5}}\colon \mathbbm{R}^{9,1|\mathbf{16}+\overline{\mathbf{16} }}\to \Omega^\infty_{B^2U(1)}\Sigma^\infty_{B^2U(1)}S^4/\!\!/ S^1
  \end{equation}
  rendering the diagram of Figure \ref{gaugeenhlift} commutative;
  
  \item the rational homotopy type of
  \begin{equation}
  \Omega^\infty_{B^2U(1)}\Sigma^\infty_{B^2U(1)}S^4/\!\!/ S^1
  \end{equation}
  contains the rational homotopy type of gerbe-twisted connective K-theory 
  \begin{equation}
  \big(\Omega^{\infty-2}_{B^2 U(1)}ku\big)/\!\!/ BU(1)
  \end{equation} as a factor; and
  
  \item the extended cocycle $\widehat{\mu_{\rm M2/M5}}$ hits precisely this twis\-ted K-theory factor, and is moreover unique up to homotopy for this specification.
\end{enumerate}
Put differently, by passing to the first-order approximation in the Goodwillie calculus of functors we obtain an unique-up-to-homotopy extension $\widehat{\mu_{\rm M2/M5}}$ of the cocycle $\widetilde{\mu_{M2/M5}}$ of \eqref{eqn:ddr1stattempt} which
\begin{enumerate}[i)]
  \item recovers the Type IIA D0, D2, F1, D4, D6 and D8-brane  cocycles\footnote{though the NS5-brane cocycle disappears---it is ``unstable'' in homotopical perturbation theory.}; and
  
  \item exhibits the combined twisted K-theory charge carried by Type IIA D-branes in the torsion-free approximation.
\end{enumerate}  
This is the main result of \cite{Braunack-Mayer:2018uyy}: the full derivation of gauge enhancement in M-theory, proceeding from M-brane charges in rational cohomotopy.

\section{Conclusion}
We conclude with some comments on how our result fits into the broader M-theory literature.
To begin with, there are all manner of twisted spectra which look like twisted K-theory in the torsion-free approximation of rational homotopy theory. Our result only applies to this rational approximation, so we have not obtained a derivation of twisted K-theory coefficients from M-theory. 
Indeed, despite wideheld beliefs within the string theory community, there is as of yet no definitive mathematically rigorous argument for twisted K-theory being the right coefficients for D-brane charge: the purported derivation of \cite{Diaconescu:2000wy} is rather a consistency check on the sign of the partition function over K-theory fields---a priori many other (twisted) cohomology theories could yield the same sign rule; the arguments of \cite{Minasian:1997mm} are differential form computations, hence cannot capture any torsion information; and there are moreover persistent conceptual issues with the proposal that D-brane charge is classified by twisted K-theory (see for example \cite{deBoer:2001wca,Evslin:2006cj}).

The main result of \cite{Braunack-Mayer:2018uyy} fills an important gap in the literature by providing the first full mathematical derivation of \emph{rational} D-brane charge from M-theory degrees of freedom.
Additionally, since we argue using universal constructions in homotopy theory, this derivation already contains hints to its own completion beyond the rational approximation.

\bibliography{allbibtex}

\bibliographystyle{prop2015}

\end{document}